\documentclass[12pt]{article}
\usepackage{amsmath}
\usepackage{amsfonts}
\def\Box{\hbox{$\sqcup$\kern-0.66em\lower0.03ex\hbox{$\sqcap$}}}

\begin{document}
\begin{titlepage}
\begin{flushright}
IFUP--TH 8/2002 \\
DFF/384/03/02 
\end{flushright}
\vskip 1truecm
\begin{center}
\Large{\bf Liouville theory, accessory parameters and $2+1$
dimensional gravity}  
\footnote{This work is  supported in part by M.I.U.R.}
\end{center}
\bigskip\bigskip
\centerline{March 2002}
\vskip 1truecm
\begin{center}
{Luigi Cantini$~^{a}$, Pietro Menotti$~^{b}$, Domenico Seminara}$^{c}$\\  
\end{center}
\begin{center}
\vskip 1truecm
\small\it $^{a}$ Scuola Normale Superiore, 56100 Pisa, Italy\\
{\small\it and INFN, Sezione di Pisa}\\
e-mail: cantini@df.unipi.it\\
{\small\it $^{b}$ Dipartimento di Fisica dell'Universit{\`a}, 56100 Pisa, 
Italy}\\
{\small\it and INFN, Sezione di Pisa}\\
e-mail: menotti@df.unipi.it\\
{\small\it $^{c}$ Dipartimento di Fisica dell'Universit{\`a}, 50125
Firenze, Italy}\\
{\small\it and INFN, Sezione di Firenze}\\
e-mail: seminara@fi.infn.it\\
\vskip 1truecm
\end{center}
\end{titlepage}

\begin{abstract} 

We prove a relation between the asymptotic behavior of the conformal
factor and the accessory parameters of the $SU(1,1)$ Riemann- Hilbert
problem. Such a relation shows the hamiltonian nature of the dynamics
of ${\cal N}$ particles coupled to $2+1$ dimensional gravity. A
generalization of such a result is used to prove a connection between
the regularized Liouville action and the accessory parameters in
presence of general elliptic singularities. This relation had been
conjectured by Polyakov in connection with $2$-dimensional quantum
gravity.  An alternative proof, which works also in presence of
parabolic singularities, is given by rewriting the regularized
Liouville action in term of a background field.

\end{abstract}

\section{Introduction}\label{Introduction}

In this paper we shall consider a problem which arises in connection
with the hamiltonian formulation of $2+1$ dimensional gravity in the
maximally slicing gauge. It is related to a variant of the
Riemann-Hilbert problem in the realm of $SU(1,1)$ monodromies and to a
conjecture by Polyakov about the auxiliary parameters occurring in the
Riemann-Hilbert problem.

From the viewpoint of 2+1 dimensional gravity coupled to point
particles one starts with the
usual ADM hamiltonian formulation  in the 
maximally slicing gauge, solves 
first the diffeomorphism and then the hamiltonian constraints and
obtains the equations of motion of particle, i.e. $\dot z_n$ and $\dot
P_n$ as function of $z_n, P_n$. The natural question is whether such
equations are canonical. As they are obtained by solving the constraint
of a canonical problem one expects and affirmative answer. However as the
expression of $\dot z_n$ and $\dot P_n$ are non trivial functions of
$z_n$ and $P_n$ a direct proof is desirable.

The proof of the hamiltonian nature of such equations was given in
ref. \cite{CMS1} in the following way. For two particle is trivial, for
three particles it involves the exploitation of the Garnier equations
related to the isomonodromic transformations. For more than three
particle the 
Garnier system of equations is not sufficient for providing the proof
of the hamiltonian nature. In \cite{CMS1} it was shown as such a result
is a consequence of an interesting conjecture due to Polyakov
\cite{conj} about
the so called accessory parameters which occur in the Riemann-Hilbert
problem.

A proof of Polyakov's conjecture was given by Zograf and
Takhtajan \cite{ZT} in the case of parabolic singularities and for
elliptic 
singularities of finite order. 2+1 dimensional gravity requires the
validity of Polyakov's conjecture for generic elliptic singularities.

Actually for proving the hamiltonian nature of the equations
of motion we shall not need the full strength of Polyakov conjecture,
but
a weak form of it will be sufficient. The weak form is of interest in
itself because it relates the constant part of the asymptotic behavior
of the reduced conformal factor to the derivative of some accessory
parameters with respect to the total energy.

Two proofs of Polyakov conjecture for general elliptic singularities
were given in \cite{CMS2} and \cite{CMS3}. Another proof was supplied 
in \cite{TZ}, where also a connection with the Kaehler metric
on moduli space has been evidenced.

Proofs \cite{CMS2} and \cite{TZ} go through a direct calculation of 
the derivative of the regularized action with respect to the position 
of the singularities. Proof \cite{CMS3} goes through an intermediate step
i.e. a weak form of Polyakov conjecture. This passage still simplifies 
the proof in the case of elliptic singularities and is based on a 
simple application of Green's formula to the linear elliptic equation
which arise when taking the derivative of Liouville equation with
respect to 
the position and strength of the singularities. By exploiting later the 
behavior of the solution as a function of the strength of the 
singularities one reaches the proof of the full Polyakov conjecture.

In this paper we shall reproduce the proof of \cite{CMS2} and \cite{CMS3}
with full details, introducing notable simplifications in them.
In this context we shall also show how the technique developed
in \cite{CMS2} immediately extends to the case of parabolic singularities.

The plan of the paper is the following: in Section 2 we summarize
the role of Liouville and uniformization theory in 2+1 dimensional
gravity
coupled to point particles.

In Section 3 we derive the canonical form of the conformal factor
in terms of the solutions of the linear second order Fuchsian
differential equation related to the uniformization problem, both
for elliptic and parabolic singularities.

In Section 4 we discuss how $SU(1,1)$ monodromies are realized in 
the context of the Riemann-Hilbert problem arising in connection with
the Liouville 
equation with point-like sources. Following such a development we prove
that for general elliptic singularities, the accessory parameters
$\beta_n$ are real analytic functions of $z_n$ and $g_n$ in an open
everywhere dense set of $R^{3{\cal N}+1}$.

In Section 5 we derive the hamiltonian nature of 2+1 dimensional 
gravity by proving a relation between the constant term in the
asymptotic behavior of the conformal factor at infinity and the accessory  
parameters $\beta_n$.

In Section 6 we use the same technique to derive a weak form
of Polyakov conjecture from which the full form easily follows.

In Section 7 give a proof of Polyakov conjecture both for elliptic 
and parabolic singularities by exploiting the technique of \cite{CMS2}
extended to the general case. The method is that to rewrite
Polyakov action in terms of a field $\phi_M$ which is less singular 
than the original $\phi$; this procedure avoids the writing of
the regularized action as a limit of an integral.

In Appendix 1 we prove the boundedness of the accessory parameters 
$\beta_n$ and of the parameter $k$ which occurs in the expression
of the conformal factor. Such result will be used in Section 4.

In Appendix 2 we give the expression of the regularized action in 
terms of the regular field $\phi_M$.

\section{2+1 dimensional gravity and Liouville theory}

The adoption of the maximally slicing gauge \cite{BCVW,MS,CMS1},
defined by  
$K=0$, where $K$ is the trace of the extrinsic curvature of the time
slice, is practically confined to the case of open universes \cite{MS}.
However in this gauge the solution of the diffeomorphism constraint 
is very simple and the hamiltonian constraint reduces to an equation 
well studied by mathematicians. The metric is parametrized in the 
standard ADM form
\begin{equation}\label{ADM}
ds^2=-N^2 dt^2+e^{2\sigma}(dz+N^zdt)(d\bar z+N^{\bar z}dt)
\end{equation}
where we have adopted the complex coordinate $z=x+iy$ and the conformal
gauge $g_{z\bar z}=\frac{1}{2} e^{2\sigma},~g_{zz} =g_{\bar z \bar z}=0$
for the space metric.

The solution of the diffeomorphism constraint is given by
\begin{equation}
\pi^{\bar z}_{~z} = -\frac{1}{2\pi}\sum_n\frac{P_n}{z-z_n},
\end{equation}
while the hamiltonian constraint reduces to
\begin{equation}\label{Hconstr}
4\partial_z \partial_{\bar z}(2\sigma)=-\pi^a_{~b}\pi^b_{~a}
e^{-2\sigma}-
\sum_n m_n\delta^2(z-z_n)
\end{equation}
being $z_n$ the particle positions and $m_n$ their rest masses.

By posing 
$\pi^a_{~b}\pi^b_{~a}e^{-2\sigma}=
2\pi^z_{~\bar z}\pi^{\bar z}_{~z}e^{-2\sigma}\equiv e^{\phi}$
eq.(\ref{Hconstr}) reduces to
\begin{equation}\label{eqphi}
4\partial_z\partial_{\bar z}\phi=e^{\phi}+4\pi \sum_n \delta^2(z- z_n)(
\mu _n -1)+4\pi\sum_B \delta^2(z- z_B),    
\end{equation}
where we have defined $\mu_n=m_n/4\pi$ and $z_B$ are the 
positions of the so called apparent singularities given by the zeros
of $\pi^{\bar z}_{~z}$ i.e.
\begin{equation}
\sum_n \frac{P_n}{z_B-z_n} = 0.
\end{equation}
Due to the restriction $\sum_n P_n=0$ \cite{MS} the $z_B$ are ${\cal N}-2$ 
in number, being ${\cal N}$ the number of particles.

As we shall see in the following section, the solution of 
eq.(\ref{eqphi}) in unique once one imposes the asymptotic behavior 
of $\phi$ at infinity $\phi = (\mu-2)\ln z\bar z + O(1)$.

The conformal factor $\phi$ plays the key role in the theory as all
other quantities can be derived algebraically from it.

The lapse and shift function are easily obtained from $\phi$
\begin{equation}
N=\frac{\partial \phi}{\partial M}
\end{equation}
being $M \equiv 4\pi \mu$ the total energy of the system and
\begin{equation}
N_z=-\frac{2}{\pi^{\bar z}_{~z}(z)}\partial_z N+g(z)
\end{equation}
where $g(z)$ is a meromorphic function which is fixed by the boundary 
condition and the absence of singularities of $N_z$ for finite $z$
\cite{MS}.

In Section 3 and 4 we shall give a general discussion of 
eq.(\ref{eqphi}), 
while we shall come back to 2+1 dimensional gravity in
Section 5.

\section{The conformal factor}

In a series of papers at the turn of the past century Picard
\cite{picard} proved that the following equation
\begin{equation}\label{pic}
4\partial_z \partial_{\bar z}\phi = e^{\phi} + 
4\pi\sum_n g_n\delta^2(z-z_n)
\end{equation}
for real $\phi$
with asymptotic behavior at infinity 
\begin{equation}
\phi(z) = -g_\infty\ln(z\bar z) + O(1)
\end{equation}
and $-1<g_n,~~1<g_\infty$ (which excludes the case of punctures) and
$\sum_n g_n +g_\infty < 0$ 
admits one and only one solution (see also \cite{poincare}). Picard
\cite{picard} achieved the solution of (\ref{pic}) through an
iteration process exploiting Schwarz alternating procedure. The same
problem has been reconsidered with variational techniques by Lichtenstein 
\cite{licht} and more recently by Troyanov \cite{troyanov}, obtaining 
results which include Picard's findings. 
The interest of such results is that they solve the following variant
of the Riemann-Hilbert problem: at $z_1,\dots z_n$ we are given not
with the monodromies but with the class, characterized by $g_j$, of
the elliptic monodromies with the further request that all such
monodromies belong to the group $SU(1,1)$. The last requirement is
imposed by the fact that the solution of eq.(\ref{pic}) has to be
single  valued. 
The inequalities
on the values of $g_m=-1+\mu_m$ and $g_\infty = 2-\mu$ 
are satisfied in $2+1$ dimensional gravity due 
to the restriction on the masses of the particles, $0<\mu_n<1$ (in
rationalized Planck units) and to the fact that the total energy $\mu$
must satisfy the bound $\sum_n \mu_n <\mu<1$. 
In this paper we shall confine ourselves to the Riemann sphere.

From eq.(\ref{pic}) one can easily prove
\cite{poincare,bilalgervais} that the function $Q(z)$ defined by
\begin{equation}\label{bg}
e^{\frac{\phi}{2}} \partial_z^2 e^{-\frac{\phi}{2}} = -Q(z)
\end{equation}
is analytic i.e. as pointed out in \cite{bilalgervais} $Q(z)$ is given
by the
analytic component of the energy momentum tensor of a Liouville
theory. $Q(z)$ is meromorphic with poles
up to the second order \cite{hempel} i.e. of the form 
\begin{equation}\label{Q}
Q(z) = \sum_n - \frac{g_n(g_n+2)}{4(z-z_n)^2} +
\frac{\beta_n}{2(z-z_n)}. 
\end{equation}
All  solutions of eq.(\ref{pic}) can be put in the form
\begin{equation}\label{mapping}
e^{\phi}=\frac{8f'\bar{f'}}{(1-f\bar f)^2} = \frac{8
|w_{12}|^2}{(y_2\bar y_2 - y_1\bar y_1)^2},~~~~
f(z) = \frac{y_1}{y_2}
\end{equation}
being $y_1,y_2$ two properly chosen, linearly independent solutions 
of the fuchsian equation 
\begin{equation}\label{fuchs}
y''+Q(z)y=0.
\end{equation}
$w_{12}$ is the constant wronskian. 
In fact following
\cite{poincare,bilalgervais} as $e^{-\phi/2}$ solves the fuchsian
equation (\ref{fuchs})
it can be put in the form
\begin{equation}\label{bgform1}
e^{-\frac{\phi}{2}}=\frac{1}{\sqrt{8}}[\psi_2(z)\bar\chi_2(\bar z) -
\psi_1(z)\bar \chi_1(\bar z)]
\end{equation}
with $\psi_j(z)$ solutions of eq.(\ref{fuchs}) with wronskian
$1$ and $\chi_j(z)$ also solutions of eq.(\ref{fuchs}) with
wronskian $1$.   
The solution of eq.(\ref{pic}) ($\phi={\rm real}$)  with the stated
behavior at infinity is unique
\cite{picard,licht,troyanov}. Exploiting the reality of $e^{\phi}$ it is
possible by an $SL(2C)$ transformation to reduce eq.(\ref{bgform1}) to
the form 
eq.(\ref{mapping}). In fact, being $\chi_j$ linear combinations of the 
$\psi_j$, the reality of $e^{\phi}$ imposes 
\begin{equation}\label{bgform2}
\psi_2(z)\bar\chi_2(\bar z) - \psi_1(z)\bar \chi_1(\bar z)=
\sum_{jk} \bar\psi_j H_{jk}\psi_k
\end{equation}  
with the $2\times 2$ matrix $H_{jk}$  hermitean and $\det H = -1$. By
means of a unitary 
transformation, which belongs to $SL(2C)$ we can reduce $H$ to
diagonal form ${\rm diag}(-\lambda, \lambda^{-1})$ and with a
subsequent $SL(2C)$ transformation we can reduce it to the form ${\rm 
diag}(-1,1)$ i.e. to the form (\ref{mapping}) where we relaxed the
condition on the wronskian $w_{12}=1$. 
Through eq.(\ref{bg}) $\phi$ contains the full information about the
accessory parameters $\beta_n$ defined in eq.(\ref{Q}). 
It is important to
notice that being all of our monodromies elliptic, we can by means of  
an $SU(1,1)$ transformation and by multiplying $y_1$ and $y_2$ by a
common factor , choose around a given singularity $z_n$ (not 
around all singularities simultaneously) $y_1$ and $y_2$ with the
following canonical behavior
\begin{equation}\label{canonical}
y_1(\zeta) = k_n \zeta^{\frac{g_n}{2}+1} A(\zeta),~~~~y_2(\zeta)
=\zeta^{-\frac{g_n}{2}}  B(\zeta)  
\end{equation}  
with $\zeta= z-z_n$ and $A$ and $B$ analytic functions of $\zeta$ in a 
neighborhood of $0$ with $A(0)=B(0)=1$.   
So the most general form for $e^{\phi}$, compatible with the reality
condition and the monodromy condition around $z_n$ is
\begin{equation}\label{phiellittico}
e^{\phi}=\frac{8 k^2_n(g_n+1)^2(\zeta \bar{\zeta})^{g_n}}{[B\bar B-
k_n^2 (\zeta \bar{\zeta})^{(g_n+1)}A\bar A]^2}
\end{equation}
and the real parameter $k_n$ is fixed by the imposition of the 
monodromy condition around the other singularities.

If we are in presence of parabolic singularities (i.e. punctures)
the Liouville equation has the form
\begin{equation}\label{liouvillepara}
4\partial_z\partial_{\bar z}\phi=e^{\phi}-4\pi
\sum_{n=1}^{N}\delta^2(z-z_n), 
\end{equation}
with asymptotic behavior at infinity 
\begin{equation}
\phi = -2\ln(z\bar z)+O(1).
\end{equation}
We recall that Picard's theorem about existence and
uniqueness 
of the solution of Liouville equation, requires the following
condition on the 
parameters $g_n, g_{\infty}$
$$
g_n>-1,~~~~g_{\infty}>1,~~~~ \sum_{n=1}^{N}g_n+g_{\infty}<0.
$$
Now $g_n=-1$, so we cannot appeal to Picard's findings in order to
assure 
existence and uniqueness of solution of eq.(\ref{liouvillepara});
however the 
problem of solving Liouville equation with parabolic singularities
is equivalent to the problem of uniformization of Riemann surfaces,
which has been  
solved by Poincar\`e and Koebe at the beginning of the last 
century \cite{poincarekoebe}.
Again one can write 
\begin{equation}
e^{-\frac{\phi}{2}}=\frac{1}{\sqrt 8}[\psi_2 \bar{\psi}_2-\psi_1
\bar{\psi}_1]. 
\end{equation}
The most important difference between the case of elliptic
and that of parabolic singularities is that, while eq.(\ref{fuchs}) in
the elliptic case leads to an indicial equation, which admits
two different solution, in the parabolic case the indicial equation is 
$$
\alpha(\alpha-1)=-\frac{1}{4}
$$
i.e. we have a double solution $\alpha=\frac{1}{2}$. In this way around 
the singularity $z_n$ we have a solution with expansion
\begin{equation}
\zeta^{\frac{1}{2}}B(\zeta)=\zeta
^{\frac{1}{2}}(1+c_2\zeta+...), 
~~~~{\rm with}~~~~\zeta=z-z_n,~~~~{\rm and}~~~~c_2= -\frac{\beta_n}{2}
\end{equation}
and a solution which contains a singularity of logarithmic type
\begin{equation}
\zeta^{\frac{1}{2}}B(\zeta)\ln(\zeta)+\zeta^{\frac{1}{2}}A_0(\zeta)
\end{equation}
where $A_0(\zeta)$ is the solution of the inhomogeneous fuchsian 
differential equation 
\begin{equation}\label{eqA}
\partial^2_{\zeta} (\zeta^{\frac{1}{2}}
A_0)
+Q(\zeta)(\zeta^{\frac{1}{2}}A_0)=\frac{1}{\zeta^2}
(\zeta^{\frac{1}{2}}B)-\frac{2}{\zeta}   
\partial_{\zeta} (\zeta^{\frac{1}{2}}B)
\end{equation}
and expansion
\begin{equation}
A_0(\zeta)=(a_1\zeta +a_2\zeta^2+a_3\zeta^3+...).
\end{equation}
In  the following we shall call
\begin{equation}
Y_1(\zeta)=\zeta^{\frac{1}{2}}B(\zeta),~~~~~
Y_2(\zeta)=-\zeta^{\frac{1}{2}}[B(\zeta)\ln \zeta +A_0(\zeta)]
\end{equation}
whose wronskian is $w_{12}=1$. In 
general a pair of linearly independent solutions of eq.(\ref{fuchs})
with wronskian equal to $1$ is
$$
\psi_1=aY_1+bY_2,~~~~\psi_2=cY_1+dY_2~~{\rm with}~~ad-bc=1.
$$
The most general form for $e^{-\frac{\phi}{2}}$ compatible with
the reality condition will be
\begin{align}
e^{-\frac{\phi}{2}}=\frac{|\zeta|}{\sqrt 8}[&(c B(\zeta)-
d B(\zeta)\ln \zeta - d A_0(\zeta))\times (c.c.)-\\&(aB(\zeta)-
b B(\zeta)\ln \zeta - b A_0(\zeta))\times (c.c)].
\end{align}
If we impose the monodromy condition around $z_n$, then we reach the 
conditions 
${\rm Im}(a\bar b-c\bar d)=0, |b|=|d|$ which together with the condition
$ad-bc=1$ imply $b=\pm\bar d$ and $a\bar b-c\bar d=\pm 1$. The
positivity of $e^{-\frac{\phi}{2}}$ chooses the solution with the
minus sign. Thus the most
general form for $e^{-\frac{\phi}{2}}$ compatible with the monodromy 
condition
around $z_n$ is
\begin{equation}\label{phiparabolico}
e^{-\frac{\phi}{2}}=-\frac{|\zeta|}{\sqrt 8}[B\bar{A}_0+ \bar
BA_0+B\bar B\ln 
\frac{\zeta \bar{\zeta}}{k_n^2}].
\end{equation}
Eq.(\ref{fuchs}) can be interpreted 
as the differential equation related to a variant of the Riemann-Hilbert
problem, in which at $z_1,...z_{\cal N}$ we are given not with the
monodromy  
but with the class of the monodromy, with the further request that all
such monodromies   
belong to the group $SU(1,1)$. The last requirement, which is
equivalent to 
the requirement that $e^{\phi}$ is a monodromic function, 
fixes in principle not only $k_n^2$ but also the values of the accessory 
parameters $\beta_n$.

\section{The realization of fuchsian SU(1,1) monodromies}

The results of Picard assure us that given the position of the 
singularities $z_n$ and the classes of monodromies characterized by
the real numbers $g_n$ and $g_\infty$ subject to the restrictions
$-1<g_n$, $1< g_\infty$ and $\sum_n g_n+g_\infty <0$,
there exists a unique fuchsian equation which realizes $SU(1,1)$
monodromies of the prescribed classes. In particular the uniqueness of
the solution of Liouville equation tells us that the accessory
parameters $\beta_i$ are single valued functions of the parameter $z_n$
and $g_n$. We shall examine in this section how such dependence arises
from the viewpoint of the imposition of the $SU(1,1)$ condition on the
monodromies in order to understand the nature of the dependence of the
$\beta_i$ on the $g_n$ and on the $z_n$.  The (non trivial) proof of
the real analytic dependence of the accessory parameters on the $z_n$
for finite order elliptic singularities has been given by Kra
\cite{kra}. 
 
Starting from the singularity in $z_1$
we can consider the canonical pair of solutions around $z_1$
i.e. those solutions which behave as a single fractional power
multiplied by an analytic function with leading coefficient $1$.
We shall call
such pair of
solutions $(y^1_1, y^1_2)$ and let $(y_1, y_2)$ the solution which
realize $SU(1,1)$ around all singularities. Obviously all conjugations
with any element of $SU(1,1)$ is still an equivalent solution in the
sense that they provide the same conformal factor $\phi$.
The canonical pairs around different singularities are linearly
related i.e. $(y^1_1, y^1_2) = (y^2_1, y^2_2) C_{21}$.
We fix the conjugation class by setting
\begin{equation}
(y_1, y_2) = (y^1_1, y^1_2) K
\end{equation}
with $K = {\rm diag}( k, 1)$ being the overall constant irrelevant
in determining $\phi$. Moreover if the solution $(y_1, y_2)$ realizes
$SU(1,1)$ monodromies around all singularities also $(y_1,
y_2)\times{\rm diag}(e^{i\alpha},e^{-i\alpha})$ accomplishes the same
purpose 
being ${\rm diag}(e^{i\alpha},e^{-i\alpha})$ an element of
$SU(1,1)$. Thus the phase of the number $k$ is irrelevant
and so we can consider it real and positive. This choice of the canonical
pairs is always possible in our case. In fact the roots of the indicial
equation are $-\frac{g_m}{2}$ and $\frac{g_m}{2}+1$
and thus the monodromy matrix has eigenvalues $e^{-i\pi g_m}$ and $e^{i\pi
g_m}$ which are different when $g_m$ is not an integer. If $g_m$ is an
integer in general in the solution of the fuchsian equation the less
singular solution possesses a logarithmic term which however has to be
absent in our case (no logarithm condition \cite{yoshida}) in order to
have a single valued $\phi$. In this case the monodromy
matrix is simply the identity or minus the identity.
The monodromy around $z_1$ thus belongs to $SU(1,1)$ for any choice of
$K$.
If $D_{n}$ denote the diagonal monodromy matrices around $z_n$, we
have that the monodromy around $z_1$ is $D_1$ and the one around $z_2$
is
\begin{equation}
M_2 = K^{-1} C_{12} D_2 C_{21} K   
\end{equation}
where with $C_{12}$ we have denoted the inverse of the $2\times 2$
matrix $C_{21}$.

In the case of three singularities (one of them at infinity) the
counting of the 
degrees of freedom is the following: by using the freedom on $K$ we
can reduce $M_2$ to the form
$\begin{pmatrix}
a & b \\ c & d
\end{pmatrix} 
$
with ${\rm Re}~ b ={\rm Re}~ c$, or if either  ${\rm Re}~ b$ or ${\rm
Re}~c$ is zero we can
obtain ${\rm Im}~ b = -{\rm Im}~ c$. Then we use the fact that $D_1 M_2
= C D_\infty C^{-1}$ and thus in addition to $a+d= {\rm
real}$ we have also $ a e^{i\pi g_1} + d e^{-i\pi g_1} =
{\rm real}$, which gives $ d= \bar a$ and thus using $a \bar a - b c
=1$ we have $c=\bar b$. The fact that a real $k$ is sufficient to
perform the described reduction of the matrix $M_2$ is assured by
Picard's result on the solubility of the problem and in this simple
case also by the explicit solution in terms of hypergeometric
functions \cite{BCVW,MS}.

We give now a qualitative discussion of the
case with four singularities and then give the analytic treatment of
it. The case with more than four singularities is a trivial extension
of the four singularity case. The following treatment relies heavily
on Picard's result about the existence and uniqueness of the solution of
eq.(\ref{pic}). 
We recall that the accessory parameters $\beta_n$ are bound by two
algebraic relations known as Fuchs relations \cite{yoshida}. Thus
after choosing
$M_1$ of the form $M_1=D_1 $,  in imposing the $SU(1,1)$ nature of
the remaining monodromies we have at our disposal three real parameters
i.e. $k$, ${\rm Re}~\beta_3$ which we shall denote by $\beta_R$ and
${\rm Im}~\beta_3$ which we shall denote by $\beta_I$. It is
sufficient
to impose the $SU(1,1)$ nature of $M_2$ and $M_3$ as the $SU(1,1)$
nature of $M_\infty$ is a consequence of them.
As the matrices $M_n= K^{-1}C_{1n}D_nC_{n1}K$ satisfy identically
$\det M_n=1$ and ${\rm Tr} M_n = 2\cos\pi g_n$ we need to impose
generically on $M_2$ only two real conditions e.g. ${\rm Re}~b_2={\rm
Re}~c_2$ and ${\rm Im}~b_2=-{\rm Im}~c_2$. The same for $M_3$. Thus it
appears that we need to satisfy four real relations when we can vary
only three real parameters.  The reason why we need only three and not
four is that for any solution of the fuchsian problem the following
relation among the monodromy matrices is identically satisfied
\begin{equation}
D_1 M_2 M_3 M_\infty=1.
\end{equation}
Rigorously the conditions for realizing $SU(1,1)$ monodromies are
\begin{equation}\label{ReIm}
{\rm Re}~a_i= {\rm Re}~d_i,~~{\rm Im}~a_i= -{\rm Im}~d_i,~~
{\rm Re}~b_i= {\rm Re}~c_i,~~{\rm Im}~b_i= -{\rm Im}~c_i~~~(i=2,3)
\end{equation}

Through a projective transformation we can bring three singularities in
$0$, $1$ and $\infty$ and we shall call $z_1$ the position of the
remaining singularity. We shall denote
the above eight relations by $\Delta^{(i)}(\beta_R,\beta_I,
k,z_1)=0$ $(i=1\dots 8)$.
Satisfying the eight above equations is a sufficient (and necessary)
condition to realize the $SU(1,1)$ monodromies.

The matrices $A_n= C_{n1}K$ which give the solution of the problem in
terms of
the canonical solutions around the singularities are completely
determined by the
two equations
\begin{equation}
(y_1,y_2) = (y^{(n)}_1,y^{(n)}_2) A_n; ~~~~
(y'_1,y'_2) = (y^{(n)'}_1,y^{(n)'}_2) A_n
\end{equation}
due to the non vanishing of the wronskian of $y^{(n)}_1,y^{(n)}_2$.
Being $(y_1,y_2)$ solutions of a fuchsian equation, $A_n$ depend
analytically  on $\beta_R, \beta_I,  z_1$.  It follows that
$\Delta^{(i)}$ are  analytic
functions of the independent variables
$\beta_R, \beta_I , k, {\rm Re} z_1,{\rm Im} z_1$. Thus
the equations $\Delta^{(i)} =0$ which
determine implicitly $k, \beta_R$ and $\beta_I$ state
the vanishing of the real analytic functions $\Delta^{(i)}$.

In Appendix 1 it is shown that for $z_1$ varying in a disk lying in
the domain $X_\epsilon$, given by the complex plane from which small
disks of radius $\epsilon$ around the singularities $z_n$ in addition
to the region $|z| >1/\epsilon$ have been removed , $\beta_R$, $\beta_I$
and $k$ are bounded function of $z_1$.
From the uniqueness of Picard's solution and the continuity of the
$\Delta^{(i)}$ it follows that the $\beta_R, \beta_I, k$ are
continuous functions of $z_1$ in $X_\epsilon$. We come now the
existence of the derivative of $\phi$ with respect to $z_n$. Due to
the real analytic dependence of $\phi$ on $\beta_R, \beta_I, k$ it is
enough to prove the existence of the derivative of such quantities
with respect to ${{\rm Re} z_1}$ and ${\rm Im} z_1$. Given a value
$z_{10}$ in $X_\epsilon$ let us consider the Picard solution
$\beta_R(z_{10})$, $\beta_I(z_{10})$, $k(z_{10})$. We have
$\Delta^{(i)}(\beta_R(z_{10}),
\beta_I(z_{10}),k(z_{10}),z_{10})=0$. The functions of $\beta_R$
$\Delta^{(i)}(\beta_R, \beta_I(z_{10}),k(z_{10}),z_{10})$ cannot all
be identically zero in a neighborhood of $\beta_R(z_{10})$ otherwise
Picard's solution would be not unique. Let be $\Delta^{(1)}(\beta_R,
\beta_I(z_{10}),k(z_{10}),z_{10})$ not identically zero in $\beta_R$
in a neighborhood of $\beta_R(z_{10})$. Then we can apply to it
Weierstrass preparation theorem \cite{gunningrossi} and write in a
neighborhood of 
$\beta_R(z_{10}), \beta_I(z_{10}),k(z_{10}), z_{10}$
\begin{equation}
\Delta^{(1)}(\beta_R, \beta_I, k, z_1) = u ~ P(\beta_R|\beta_I, k,
{\rm Re}(z_{1}),{\rm Im}(z_{1}))
\end{equation} 
where $u$ is a unit and $P$ is a monic polynomial in $\beta_R$ and
coefficients analytic functions of the remaining variables. In the
Weierstrass neighborhood all the zeros of $\Delta^{(1)}$ are zeros of
$P$ and 
viceversa.
In addition we have $P(\beta_R(z_{10})|\beta_I(z_{10}), k(z_{10}), {\rm
Re}(z_{10}),{\rm Im}(z_{10}))=0$.  

\smallskip

a) If $P'(\beta_R(z_{10})|\beta_I(z_{10}), k(z_{10}), {\rm Re}(z_{10}),{\rm
Im}(z_{10}))\neq 0 $ we can solve for $\beta_R$ and obtain
$\beta_R = \beta_R (\beta_I, k, {\rm Re}(z_{1}),{\rm Im}(z_{1}))$ in
the Weierstrass neighborhood of $\beta_R(z_{10}),
\beta_I(z_{10}),k(z_{10}), z_{10}$.

\smallskip

b) If $P'(\beta_R(z_{10})|\beta_I(z_{10}), k(z_{10}), {\rm
Re}(z_{10}),{\rm Im}(z_{10}))=0$ but
$P'(\beta_R(z_{1})|\beta_I(z_{1}), k(z_{1}), {\rm
Re}(z_{1}),\break {\rm Im}(z_{1}))$ not identically zero in a
neighborhood of $z_{10}$ then we can solve $\beta_R = \break \beta_R
(\beta_I, k, {\rm Re}(z_{1}),{\rm Im}(z_{1}))$ in a dense open set
around 
$z_{10}$.

\smallskip

c) If $P'(\beta_R(z_{1})|\beta_I(z_{1}), k(z_{1}), {\rm
Re}(z_{1}),{\rm Im}(z_{1}))$ is identically zero in a neighborhood of
$z_{10}$ then we consider $P''(\beta_R|\beta_I, k, {\rm
Re}(z_{1}),{\rm Im}(z_{1}))$ and proceed as above. 

\smallskip 

Being the Weierstrass polynomial monic the process in c) ends in a
finite 
number of steps with the result that we are able to solve
$\beta_R$ as an analytic function of $\beta_I$, $k$ and $z_1$ for
$z_1$ in a dense open set around $z_{10}$.

One can substitute  the obtained $\beta_R$ in the $\Delta^{(i)}$
which become analytic functions of only $\beta_I, k, {\rm 
Re}(z_{1}),{\rm Im}(z_{1})$. Then one eliminates by the same
procedure $\beta_I$ and then $k$ with the final result that $\beta_R$,
$\beta_I$ and $k$ are real analytic functions of ${\rm Re}(z_{1}),{\rm
Im}(z_{1})$ in a everywhere dense open set of $R^2$. 

The procedure is
immediately extended to any number of singularities and also to the
case of parabolic singularities.  Our result is not as
strong as the one given by Kra \cite{kra} in the special 
case of elliptic singularities of finite order and the one 
given by Zograf and Takhtajan \cite{ZT} in the case of parabolic
singularities. 

Similarly one can deal with the dependence of the $\beta_n$ and $k^2$ 
as functions of the $g_n$.

\section{Proof of the hamiltonian nature of 2+1 dimensional gravity} 

In this section we shall analyze the asymptotic behavior of the
solution of the Liouville equation (\ref{pic}) in the case of elliptic
singularities, proving several
useful relations, in particular  we shall prove that the derivative of
the 
constant term of the expansion of $\phi$ at infinity with respect of
the position of a singularity $z_n$ equals the value of the derivative
of 
the auxiliary parameters $\beta_n$ related with the same singularity
with respect to the parameter $g_\infty$.

First of all we examine the expansion of  $\phi$  around the
singularity $z_n$ 
\begin{equation}
\phi(z) = g_n \ln(z-z_n)(\bar z-\bar z_n) + r_n(z)
\end{equation}
being $r_n(z)$ a continuous function in a finite neighborhood of $z_n$
and at infinity
\begin{equation}
\phi(\frac{1}{w}) = -g_\infty \ln(\frac{1}{w \bar w})+ r_\infty(w)
\end{equation}
being $r_\infty(w)$ a continuous function in a finite neighborhood of
$w=0$. 

To make explicit the form of  $r_n(z)$ we can compute the coefficients 
in the expansions $A=1+c_1\zeta+O(\zeta^2)$ and
$B=1+c_2\zeta+O(\zeta^2)$, with $\zeta = z - z_n$, 
which are known from the fuchsian differential equation
\begin{equation}
c_1=-\frac{\beta_n}{2(2+g_n)}~~~~{\rm
and}~~~~c_2=\frac{\beta_n}{2g_n}.
\end{equation}

Then we can substitute the expansion of $A$ and $B$ into
eq.(\ref{phiellittico}), where  
$k_n$ is fixed by the global requirement of the $SU(1,1)$ nature of
the monodromies.  

By taking the logarithm of (\ref{phiellittico}) we have
$$
\phi =g_n\ln\zeta\bar\zeta + \ln8 k^2_n(g_n+1)^2 -
2\ln\left(|1+c_2\zeta 
+ \cdots|^2 - k_n^2 (\zeta\bar\zeta)^{g_n+1}
|1+c_1\zeta+\cdots|^2\right) = 
$$
\begin{equation}\label{expansion}
=g_n\ln\zeta\bar\zeta -\ln s_n^2 - 2(c_2\zeta+\overline{
c_2}\bar\zeta) +O(|\zeta|^2) +
\end{equation}
$$
+2 k_n^2(\zeta\bar\zeta)^{g_n+1}(1+O(|\zeta|))+O(|\zeta|^{4(g_n+1)})
$$
where we designed by $-\ln s_n^2$ the constant term in the expansion
and $s_n^2  =1/8 k^2_n(g_n+1)^2$. 

Similarly at infinity we have
$$
\phi =-g_{\infty}\ln z \bar z -\ln s^2_{\infty} -
2\ln\left(|1+\frac{c_2}{z} 
+ \cdots|^2 - k_\infty^2 ( z \bar z )^{1-g_{\infty}}
|1+\frac{c_1}{ z}+\cdots|^2\right) = 
$$
\begin{equation}
=-g_{\infty}\ln z \bar z -\ln s^2_{\infty} -
2(\frac{c_2}{z}+\frac{\overline{c_2}}{\bar z }) +O(|\frac{1}{z}|^2) +
\end{equation}
$$
+2 k_\infty^2( z \bar z)^{1-g_{\infty}}\left(1+O(\frac{1}{|z|}) \right) + 
O((z\bar z)^{2-2g_{\infty}}).
$$

We shall now prove the following result
\begin{equation}
\frac{\partial \ln s^2_{\infty}}{\partial z_n} = \frac{\partial
\beta_n}{\partial  g_\infty}.
\end{equation}
In order to do so let us consider an elliptic equation of the form 
\begin{equation}\label{semilinear}
\partial_z\partial_{\bar z} \phi = F(\phi)
\end{equation}
and let $\phi(z,v_1, \dots v_K)$ be a family of solutions of
eq.(\ref{semilinear})  depending on the parameters
$v_1, \dots v_K$ in some given domain $D$ of the complex plane. By
taking the derivative of eq.(\ref{semilinear}) with respect to $v_i$
we have
\begin{equation}\label{linear}
\partial_z \partial_{\bar z} \frac{\partial \phi}{\partial v_i} =
F'(\phi)\frac{\partial \phi}{\partial v_i}. 
\end{equation} 
Then from eq.(\ref{linear}) we have in $D$
\begin{equation}\label{green}
\partial_z\left(\frac{\partial \phi}{\partial v_j}\partial_{\bar z}
(\frac{\partial \phi}{\partial v_i})\right)-
\partial_{\bar z}\left(\frac{\partial \phi}{\partial v_i}
\partial_z(\frac{\partial \phi}{\partial v_j})\right)=0.
\end{equation} 
We shall now apply eq.(\ref{green}) to the inhomogeneous Liouville
equation  
\begin{equation}\label{liouville}
4\partial_z\partial_{\bar z} \phi = e^{\phi} +4\pi\sum_n
g_n\delta^2(z-z_n) 
\end{equation}
in a domain which excludes the sources. 
The solutions of eq.(\ref{liouville}) depend on the parameters
$z_n, g_n, 
g_\infty$ which now play the role of the parameters $v_i$. We apply
eq.(\ref{green}) choosing as first parameter $g_\infty$ and second
parameter $z_n$, specifying the  domain $X_\epsilon$ as a disk of
radius $R$  
which includes all singularities, from which disks of radius
$\epsilon$ have been removed. Using Stokes' theorem we obtain 
$$
0=\oint_{R}  \left( \frac{\partial \phi}{\partial z_n } 
\partial_z(\frac{\partial \phi}{\partial g_{\infty}})\right)\frac{i}{2}dz  
+\left( \frac{\partial \phi}{\partial  g_{\infty}} 
\partial_{\bar z}(\frac{\partial \phi}{\partial
z_n})\right)\frac{i}{2}d\bar z   
$$
\begin{equation}\label{stokes}
-\sum_l \oint_{\gamma_l} 
\left( \frac{\partial \phi}{\partial z_n } 
\partial_z(\frac{\partial \phi}{\partial g_{\infty}})\right)\frac{i}{2}dz  
+\left( \frac{\partial \phi}{\partial  g_{\infty}} 
\partial_{\bar z}(\frac{\partial \phi}{\partial
z_n})\right)\frac{i}{2}d\bar z.   
\end{equation}
Let us consider the first integral; the first term gives in the limit
$R\rightarrow \infty$  
\begin{equation}
-\pi \frac{\partial \ln s^2_{\infty}}{\partial z_n}
\end{equation}
while the second
\begin{equation}
R \ln R^2 ~O(\frac{1}{R^2})\rightarrow 0.
\end{equation}

We consider now the contribution of the integral around the circle of
center $z_n$ and radius $\epsilon_n$; we have for the first term
\begin{equation}
-\oint_{\gamma_n} 
\left( \frac{\partial \phi}{\partial z_n } 
\partial_z(\frac{\partial \phi}{\partial
g_{\infty}})\right)\frac{i}{2}dz = 
\end{equation}
$$ 
-\oint \frac{i}{2}d\zeta \left(\frac{-g_n}{\zeta}
+\frac{\beta_n}{g_n}-\frac{\partial \ln s^2_n}{\partial z_n}+O(|\zeta|)
+ \frac{O(|\zeta\bar \zeta|^{g_n+1})}{\zeta}\right)\times
$$
$$
\left(-\frac{1}{g_n}\frac{\partial\beta_n}
{\partial g_{\infty}} + 2 (g_n+1) \frac{\partial k_n^2}{\partial
g_\infty} \frac{(\zeta\bar\zeta)^{g_n+1}}{\zeta}+\dots
\right) \rightarrow \pi \frac{\partial\beta_n}{\partial g_{\infty}}
$$
where the terms like
\begin{equation}
\oint \frac{i}{2}d\zeta(\frac{-g_n}{\zeta})\times 2 (g_n+1)
\frac{\partial k_n^2}{\partial 
g_\infty} \frac{(\zeta\bar\zeta)^{g_n+1}}{\zeta} 
\end{equation}
which do not vanish by power counting are identically zero by the
phase integration.
 
Similarly
\begin{equation}
\oint_{\gamma_n}( \frac{\partial \phi}{\partial  g_{\infty}} 
\partial_{\bar z}(\frac{\partial \phi}{\partial z_n}))\frac{i}{2}d\bar
z =  
\end{equation}
$$
\oint \frac{i}{2}d\bar {\zeta} 
\left(-\frac{\partial \ln s^2_n}{\partial g_{\infty}} -\frac{1}{g_n}
\frac{\partial \beta_n}{\partial g_{\infty}}\zeta -
\frac{1}{g_n}
\frac{\partial \bar\beta_n}{\partial g_{\infty}}\bar\zeta
+ \dots \right)
\left(-2\frac{\partial \bar c_2}{\partial z_n}+ \dots 
\right)
\rightarrow 0
$$
thus obtaining
\begin{equation}\label{crossderivatives}
\frac{\partial \ln s^2_{\infty}}{\partial z_n} = \frac{\partial\beta_n}
{\partial g_{\infty}}.
\end{equation} 

Relation (\ref{crossderivatives}) is of  fundamental importance in 
canonical 2+1 dimensional gravity; in fact the hamiltonian nature of  
particles dynamics can be proved by means of this relation, 
and the hamiltonian is just $\ln s^2_{\infty}$ .
In fact we recall that the equations of motions in the rotating frame
are (see \cite{CMS1})
\begin{equation}\label{dotz1}
\dot z'_n = -\sum_B\frac{\partial\beta_B}{\partial\mu}\frac{\partial
z'_B}{\partial P'_n} ~~~~~~~~n=2,\dots {\cal N} 
\end{equation}
and
\begin{equation}\label{dotP1}
\dot P'_n = \frac{\partial\beta_n}{\partial\mu}+
\sum_B\frac{\partial\beta_B}{\partial\mu}\frac{\partial z'_B}{\partial
z'_n}  ~~~~~~~~n=2,\dots {\cal N}, 
\end{equation}
where the indices $B$ label the so called apparent singularities,
$n$ label the particles and $z'_n= z_n-z_1$ and $P'_n = P_n$
$(n=2,\dots,{\cal N})$, 
The previous equations take the form
\begin{equation}\label{dotz11}
\dot z'_n = \sum_B \frac{\partial \ln s^2_{\infty}}{\partial z_B}
\frac{\partial 
z'_B}{\partial P'_n}= \frac{\partial \ln s^2_{\infty}}{\partial P'_n}
~~~~~~~~n=2,\dots {\cal N} 
\end{equation}
and
\begin{equation}\label{dotP11}
\dot P'_n = -\frac{\partial \ln s^2_{\infty}}{\partial z'_n}|_{z'_B} -
\sum_B\frac{\partial \ln s^2_{\infty}}{\partial z'_B} \frac{\partial
z'_B}{\partial 
z'_n}  = -\frac{\partial \ln s^2_{\infty}}{\partial z'_n}
~~~~~~~~n=2,\dots {\cal N}. 
\end{equation}
So $\ln s^2_{\infty}$ is the hamiltonian for the ${\cal N}$ particle
system. As $\ln s^2_{\infty}$ is a single valued function of the
arguments 
we have that the hamiltonian is a single valued function on the
phase space and not a section.

\section{Proof of Polyakov conjecture for general elliptic
singularities}

In this section we shall give a proof of Polyakov conjecture for
general elliptic singularities following the path outlined in the
previous paragraph, generalizing relation (\ref{crossderivatives}) to
all $g_n$ and $\ln s^2_m$. This results in a weak form of Polyakov
relation. From the weak form the strong form is immediately
obtained. This method provides the simplest proof of
Polyakov conjecture for elliptic singularities.

We exploit the relations that can be found applying
eq.(\ref{stokes}) choosing the couples of  parameters $(z_n,
g_m)$, $(g_n, g_m)$, $(z_n, z_m)$. 
Let us start with $(z_n, g_m)$ (and $n \neq m$). The only surviving
contributions are those at $z_n$ and $z_m$, the first giving
\begin{equation}
\oint_{\gamma_n} 
\left( \frac{\partial \phi}{\partial z_n } 
\partial_z(\frac{\partial \phi}{\partial g_m})\right)\frac{i}{2}dz  
\end{equation} 
$$
= \oint_{\gamma} \frac{i}{2}d\zeta~ \left(-\frac{g_n}{\zeta}+ \dots \right)
(-\frac{1}{g_n}\frac{\partial\beta_n}
{\partial g_m}+\dots)
\rightarrow -\pi\frac{\partial\beta_n}{\partial g_m}
$$
and
\begin{equation}
\oint_{\gamma_n}  
\left( \frac{\partial \phi}{\partial  g_m} 
\partial_{\bar z}(\frac{\partial \phi}{\partial
z_n})\right)\frac{i}{2}d\bar z \rightarrow 0.
\end{equation}
Around the circle $z_m$ the integration gives
\begin{equation}
\oint_{\gamma_m} 
\left( \frac{\partial \phi}{\partial z_n } 
\partial_z(\frac{\partial \phi}{\partial g_m})\right)\frac{i}{2}dz  
\end{equation} 
$$
= \oint_{\gamma} \frac{i}{2}dz~ \left(
-\frac{\partial \ln s_m^2}{\partial z_n }+ \dots \right)
\left(\frac{1}{\zeta}+  \dots \right)
\rightarrow \pi \frac{\partial \ln s_m^2}{\partial z_n}
$$
while
\begin{equation}
\oint_{R}  
\left( \frac{\partial \phi}{\partial  g_m} 
\partial_{\bar z}(\frac{\partial \phi}{\partial
z_n})\right)\frac{i}{2}d\bar z   
\rightarrow 0.
\end{equation}
In this way we have reached 
\begin{equation}\label{dbetadg}
\frac{\partial \ln s_m^2}{\partial
z_n}=\frac{\partial\beta_n}{\partial g_m}.  
\end{equation}
For $n=m$ both contributions come from the circle of center $z_m$ but
the result is the same, that is
\begin{equation}
\frac{\partial \ln s_m^2}{\partial
z_m}=\frac{\partial\beta_m}{\partial g_m}.  
\end{equation}
When in eq.(\ref{stokes}) we take the derivative with respect to
$g_m$ and $g_n$ of $\phi$ we reach
\begin{equation}
\frac{\partial \ln s^2_m}{\partial g_n} =\frac{\partial \ln
s^2_n}{\partial 
g_m},
\end{equation}
and when the chosen parameters are $z_n$ and $z_m$ we reach the 
relation
\begin{equation}
\frac{\partial \beta_m}{\partial z_n} =\frac{\partial \beta_n}{\partial
z_m}.
\end{equation}
We can summarize the previous results by stating that the form $\omega$
defined by
\begin{equation}
\omega=\sum_n \beta_n dz_n + \sum_n\ln s^2_ndg_n + c.c.
\end{equation}
is closed.
Finally we observe that even the derivative 
$\frac{\partial \phi}{\partial z}$
satisfies in $D$ the linear equation 
\begin{equation}
4\partial_z\partial_{\bar z} \frac{\partial \phi}{\partial z} = 
F'(\phi)\frac{\partial
\phi}{\partial z} 
\end{equation}
but if we study the relation
$$
0=\oint_{R}  \left( \frac{\partial \phi}{\partial z } 
\partial_z(\frac{\partial \phi}{\partial v})\right)\frac{i}{2}dz  
+\left( \frac{\partial \phi}{\partial  v} 
\partial_{\bar z}(\frac{\partial \phi}{\partial
z})\right)\frac{i}{2}d\bar z  
$$
\begin{equation}
-\sum_l \oint_{\gamma_l} 
\left( \frac{\partial \phi}{\partial z } 
\partial_z(\frac{\partial \phi}{\partial v})\right)\frac{i}{2}dz  
+\left( \frac{\partial \phi}{\partial  v} 
\partial_{\bar z}(\frac{\partial \phi}{\partial
z})\right)\frac{i}{2}d\bar z   
\end{equation}
with $v=g_{\infty}, g_n, z_n$ we 
find simply the vanishing of the derivative of the well-known first
Fuchs relation $\sum_n \beta_n = 0$, with respect to $v$.

We shall now relate the previous results to the regularized
Liouville action \cite{takh1}
\begin{equation}
S_P[\phi]=\lim_{\epsilon \rightarrow 0} S_{\epsilon}[\phi]
\end{equation}
where  
$$
S_\epsilon [\phi] =\frac{i}{2} \int_{X_\epsilon} (\partial_z\phi 
\partial_{\bar z} \phi +\frac{e^\phi}{2}) dz\wedge d\bar z
+\frac{i}{2}\sum_n g_n\oint_n\phi(\frac{d\bar z}{\bar z -\bar
z_n}- \frac{d z}{ z - z_n})
$$
\begin{equation}\label{Sepsilon1}
+\frac{i}{2}g_\infty\oint_\infty\phi(\frac{d\bar z}{\bar z}- \frac{d
z}{z})
-\pi\sum_n g_n^2 \ln\epsilon^2 -\pi g_\infty^2\ln\epsilon^2. 
\end{equation}
$X_\epsilon$ is a disk of radius $1/\epsilon$  
which includes all singularities, from which disks of radius
$\epsilon$ around each singularity have been removed.
The variation of such action provides the inhomogeneous Liouville
equation (\ref{pic}). The contour terms impose the correct behavior of
the $\phi$ on the singularities and at infinity. We shall now compute
the derivative with respect to $g_m$ of $S_\epsilon$ calculated on the
solutions of eq.(\ref{pic}). As the action is stationary on the solution
of eq.(\ref{pic}) the only contribution is due the term in
eq.(\ref{Sepsilon1}) 
which explicitly depend on $g_m$. A simple computation gives
\begin{equation}\label{dSdg}
\frac{\partial S_P}{\partial g_m} = -2\pi \ln s_m^2.
\end{equation}    
Putting together the results of eqs.(\ref{dbetadg},\ref{dSdg}) we have 
\begin{equation}
-\frac{1}{2\pi}d \frac{\partial S_P[\phi]}{\partial g_m }= 
\sum_n \frac{\partial \beta_n}{\partial g_m } dz_n+ c.c.
\end{equation}
from which
\begin{equation}
-\frac{1}{2\pi}d S_P[\phi]= 
\sum_n \beta_n dz_n+ c.c. + F
\end{equation}
where
\begin{equation}
F = \sum_n f_n dz_n +c.c.
\end{equation}
with $f_n$ dependent on $z_n$ but not on $g_n$.

We prove now that $F=0$. To this end we shall take one of the
parameters $g_n$ e.g. $g_1$ to zero. Recalling the expansion of $\phi$
around $z_n$ 
\begin{equation}
\phi = g_1\ln\zeta\bar\zeta +O(1) -2  c_2 \zeta -2  \bar{c_2}
\bar\zeta 
+ 2 k_1^2 (\zeta\bar\zeta)^{g_1+1}+\cdots
\end{equation}
we can compute $Q(z)$
\begin{equation}
Q(z) = \frac{1}{2}\partial_z^2 \phi -\frac{1}{4}(\partial_z \phi)^2 =
-\frac{g_1(g_1+2)}{4\zeta^2} + \frac{g_1 c_2}{\zeta}+ \cdots
\end{equation}
from which $\beta_1 = 2 g_1 c_2$. For $g_1\rightarrow 0$ both terms
disappear from $Q(z)$ and $g_1$ disappears from the r.h.s. of
eq.(\ref{pic}) as well (cfr. also \cite{hempel}). Now $S_P[\phi]$
depends on $z_2,\cdots z_N$ and 
\begin{equation}
-\frac{1}{2\pi}dS_P[\phi] = \beta_2 dz_2+ \cdots +\beta_N dz_N + f_1
dz_1 +f_2 dz_2 +\cdots f_N dz_N + c.c.
\end{equation}
from which it follows $f_1\equiv 0$. Similarly we reason with the
other singularities obtaining $F=0$.

\section{Direct proof, elliptic and parabolic case}\label{direct}

The previous proof goes through the weak form of Polyakov relation
as an intermediate step. We think this is the simplest way to prove
Polyakov conjecture. On the other hand, as one has to take the 
derivative with respect to $g_m$, this path cannot be followed in 
the case of parabolic singularities.
In \cite{CMS2} we gave a direct proof of Polyakov conjecture by rewriting
the action in term of some background fields. In the following we shall 
follow this path introducing some short cuts with respect to the proof in
\cite{CMS2} (see also \cite{TZ} for a different approach). 
Even if the treatment is not as simple as the one given in Section 4
it has the advantage of being immediately extendible to the case of 
parabolic singularities. The case of parabolic singularities was first
solved in \cite{ZT}; on the other hand we think worth while to see how the 
present method applies to both cases.

The technique to 
prove Polyakov conjecture will be to express the 
original action in terms of a 
field $\phi_M$ which is less singular than the original conformal
field $\phi$. This procedure will  
give rise to an action $S$ for the field $\phi_M$ which does not
involve the $\epsilon\rightarrow 0$ process.
Despite that, computing the derivative of the new action $S$ 
is not completely trivial because one cannot take directly the
derivative operation under the integral sign. In fact such unwarranted
procedure would give rise to an integrand which is not absolutely
summable. This does not apply to the simpler case of the derivative
with respect to $g_m$ considered in the previous section, as in that
case the
derivative of the integrand is absolutely summable and bounded by a
summable function as $g_m$ varies in an interval. 

In the global coordinate system $z$ on $\mathbb{C}$  one writes 
$\phi = \phi_M +\phi_1+\alpha_1\phi_B$  
where $\phi_B$ is a background conformal factor which is regular and
behaves at infinity  like $\phi_B = -2\ln(z\bar z)+c_B+O(1/|z|)$ 
(a possible choice for $\phi_B$ is the conformal factor of the sphere
with constant curvature $e^{\phi_B} = \frac{8}{(1+z\bar z)^2}$) 
while $\phi_1$ is defined by
\begin{equation}
\phi_1 = \sum_n g_n \ln|z-z_n|^2 + c_0.
\end{equation}
Then
we have for $\phi_M$
\begin{equation}\label{eqphiM}
4\partial_z\partial_{\bar z}\phi_M =
e^{\phi_M+\phi_1+\alpha_1 \phi_B}- \alpha_14~\partial_z\partial_{\bar
z} \phi_B.    
\end{equation}
We shall choose $\alpha_1 = (\sum_n g_n+g_\infty)/2$ so that $\phi_M$
will be a function regular at infinity.
The action which generates the above equation is
\begin{equation}\label{Saction}
S=
\int \left(
\partial_z\phi_M \partial_{\bar z}\phi_M + \frac{e^{\phi}}{2}
-2\alpha_1\phi_M\partial_z\partial_{\bar z}\phi_B\right) \frac{i
dz\wedge 
d\bar z}{2}. 
\end{equation}
The fields $\phi_1$ and $\phi_M$ transform under a change of chart
like scalars while $e^{\phi_B}$ transforms as a $(1,1)$ density. This
choice is also in agreement with the invariance of
eqs.(\ref{eqphiM},\ref{Saction}). 
Due to the behavior of
$\phi_M$ and $\phi_1$ at the singularities and at infinity the
integral in eq.(\ref{Saction}) converges absolutely. 

The regularized Liouville action is given by \cite{takh1}
\begin{equation}
S_P[\phi]=\lim_{\epsilon \rightarrow 0} S_{\epsilon}[\phi]
\end{equation}
where
$$
S_\epsilon [\phi] =\frac{i}{2} \int_{X_\epsilon} (\partial_z\phi 
\partial_{\bar z} \phi +\frac{e^\phi}{2}) dz\wedge d\bar z
+\frac{i}{2}\sum_n g_n\oint_n\phi(\frac{d\bar z}{\bar z -\bar
z_n}- \frac{d z}{ z - z_n})
$$
\begin{equation}\label{Sepsilon}
+\frac{i}{2}g_\infty\oint_\infty\phi(\frac{d\bar z}{\bar z}- \frac{d
z}{z})
-\pi\sum_n g_n^2 \ln\epsilon^2 -\pi g_\infty^2\ln\epsilon^2 
\end{equation}
both for elliptic and parabolic singularities (in which case $g_n=-1,~
g_{\infty}=2$), even if the behavior of the solutions of the
Liouville equation  
around parabolic singularities is completely different from that 
around elliptic singularities. 
It is proven in Appendix 2 that the action $S$ computed on a function
$\phi$, with the following asymptotic behavior at the singularities
\begin{displaymath}
\left\{
\begin{array}{ll}
g_m \ln \zeta \bar {\zeta}& \textrm{around an elliptic singularity}\\
-\ln (\zeta \bar {\zeta}) -\ln (\ln (\zeta \bar {\zeta}))^2&
\textrm{around a parabolic singularity}\\
-g_{\infty}\ln z \bar z& \textrm{at infinity}
\end{array} \right.
\end{displaymath}
 is
related to the original Liouville action $S_P$ by 
$$
S_P = S +\pi\sum_m\sum_{n\neq m} g_m g_n \ln|z_m-z_n|^2+
$$
\begin{equation}\label{SPtoS}
4\pi \alpha_1 c_0 -\alpha_1^2\int \phi_B\partial_z\partial_{\bar z}\phi_B
\frac{idz\wedge d\bar z}{2} + 2\pi \alpha_1^2c_B. 
\end{equation}

We saw in Section 4 how on an open everywhere dense set of $\mathbb{C}$
there 
exists the derivative with respect to $z_n$ of the parameters $k,{\rm
Re}\beta_i,{\rm Im}\beta_i$ which determine the solutions of the
fuchsian equation related by eq.(\ref{mapping}) to the conformal factor
$\phi$. Actually as pointed out at the end of Section 4 in that domain 
such parameters are real analytic functions of $z_n$. On the
other hand the solutions of the fuchsian equation and 
thus $\phi_M$ depend analytically on such parameters \cite{hille}.

The procedure to compute the derivative will be to prove that 
\begin{equation}\label{limit}
\frac{\partial S}{\partial z_m}= \lim_{\epsilon \rightarrow 0}
\int_{X_\epsilon} \frac{\partial F}{\partial z_m} 
\frac{i}{2}dz\wedge d\bar z
\end{equation}
where $X_\epsilon$ has been defined after eq.(\ref{Sepsilon1})
and $F$ is given by
\begin{equation}\label{defF}
F= \partial_z \phi_M
\partial_{\bar z}\phi_M 
+\frac{e^\phi}{2}
-2 \alpha_1\phi_M\partial_z
\partial_{\bar z}\phi_B.
\end{equation}
First we write the identity
\begin{equation}\label{decompS}
S=
\int_{X_\epsilon}
F~ \frac{i}{2}dz\wedge d\bar z+
\int_{\mathbb{C}\setminus X_\epsilon}
F~ \frac{i}{2}dz\wedge d\bar z.
\end{equation}
The second integral is a sum of integrals having as domain
disks of radius $\epsilon$ around each singularity.
If we take the derivative with respect to $z_m$ we have
\begin{equation}\label{deriv1}
\frac{\partial S}{\partial z_m}=
\frac{\partial}{\partial z_m}
\int_{X_\epsilon}
F~ \frac{i}{2}dz\wedge d\bar z+
\frac{\partial}{\partial z_m}
\int_{\mathbb{C} \setminus X_\epsilon}
F~ \frac{i}{2}dz\wedge d\bar z,
\end{equation}
and the second term on the r.h.s. goes to zero as $\epsilon
\rightarrow 0$. It will be sufficient to prove this for the contribution
coming from the disk $D^m_\epsilon$ around the singularity $z_m$; in
the following we shall denote with $D^0_\epsilon$ the disk of center
$0$ and radius $\epsilon$. In
the elliptic case  
we have the structure
$$
\frac{\partial}{\partial z_m}
\int_{D^m_\epsilon}
F~ \frac{i}{2}dz\wedge d\bar z=
$$
$$
\frac{\partial}{\partial z_m}
\int_{D^0_\epsilon}
\left( \frac{2k^2_m(1+g_m)A\bar A(\zeta \bar{\zeta})^{1+g_m}}
{\zeta[B\bar B -k_m^2A\bar A(\zeta \bar{\zeta})^{1+g_m}]}
+\dots \right)
\left( \frac{2k^2_m(1+g_m)A\bar A(\zeta \bar{\zeta})^{1+g_m}}
{\bar{\zeta}[B\bar B -k_m^2A\bar A(\zeta \bar{\zeta})^{1+g_m}]}
+\dots \right)\frac{i}{2} d\zeta \wedge d\bar {\zeta}+
$$
$$
\frac{\partial}{\partial z_m}
\int_{D^0_\epsilon}
\frac{4k^2_m(g_m+1)^2(\zeta \bar{\zeta})^{g_m}}{[B\bar B-
k_m^2 (\zeta \bar{\zeta})^{(g_m+1)}A\bar A]^2}
\frac{i}{2} d\zeta \wedge d\bar {\zeta}+
$$
\begin{equation}
\frac{\partial}{\partial z_m}
2\alpha_1
\int_{D^0_\epsilon}
\left(\ln [B\bar B -k_m^2 A\bar A (\zeta
\bar{\zeta})^{1+g_m}]^2+\dots\right) 
\partial_z 
\partial_{\bar z}\phi_B(\zeta+z_m)
\frac{i}{2} d\zeta \wedge d\bar {\zeta},
\end{equation}
and in the parabolic case
$$
\frac{\partial}{\partial z_m}\int_{D^0_\epsilon}
\left(\frac{-2}{\zeta [\frac{\bar {A_0}}{\bar B}+\frac{A_0}{B}+
\ln\frac{\zeta \bar {\zeta}}{k_m^2}]}+\dots\right)
\left(\frac{-2}{\bar {\zeta} [\frac{\bar {A_0}}{\bar B}+\frac{A_0}{B}+
\ln\frac{\zeta \bar {\zeta}}{k_m^2}]}+\dots\right)
\frac{i}{2} d\zeta \wedge d\bar {\zeta}+
$$
$$
\frac{\partial}{\partial z_m}\int_{D^0_\epsilon}
\frac{4}{\zeta \bar {\zeta}B\bar B
[\frac{\bar {A_0}}{\bar B}+\frac{A_0}{B}+
\ln\frac{\zeta \bar {\zeta}}{k_m^2}]^2}
\frac{i}{2} d\zeta \wedge d\bar {\zeta}+
$$
\begin{equation}
2\alpha_1\frac{\partial}{\partial z_m}\int_{D^0_\epsilon}
\left(\ln[\frac{\bar {A_0}}{\bar B}+\frac{A_0}{B}+
\ln\frac{\zeta \bar {\zeta}}{k_m^2}]^2+\dots\right)\partial_z\partial_
{\bar z}\phi_B
\frac{i}{2} d\zeta \wedge d\bar {\zeta}.
\end{equation}
It is possible to commute derivative and integral because the
derivative 
of the integrand is an absolutely summable function, bounded by an
absolutely 
summable function independent of $z_m$, when $z_m$ varies in a 
small interval.
So, being the integrand absolutely summable, if we take the limit in
which
the domain of integration vanishes we obtain zero.

We come now to the first term of eq.(\ref{deriv1}).
It is possible to rewrite it as
\begin{equation}
\frac{\partial}{\partial z_m}
\int_{X_\epsilon}
F~ \frac{i}{2}dz\wedge d\bar z=
\int_{X_\epsilon}
\frac{\partial F}{\partial z_m} \frac{i}{2}dz\wedge d\bar z -
\oint_{\partial D^m_\epsilon}
F~ \frac{i}{2} d\bar z 
\end{equation}
where the contour term comes from the movement of the 
domain of integration. In the limit $\epsilon \rightarrow 0$
this last term vanishes, so we are left with
\begin{equation}
\frac{\partial S}{\partial z_m}=
\lim_{\epsilon \rightarrow 0}
\int_{X_\epsilon}
\frac{\partial}{\partial z_m}
\left(\partial_z \phi_M
\partial_{\bar z}\phi_M +\frac{e^\phi}{2}- 2 \alpha_1\phi_M\partial_z
\partial_{\bar z}\phi_B\right)\frac{i}{2}dz\wedge d\bar z.
\end{equation}
Using now the equation of motion (\ref{eqphiM}) we obtain 
\begin{equation}
\frac{\partial S}{\partial z_m} = \lim_{\epsilon
\rightarrow 0} 
\int_{X_\epsilon}\left(\partial_z(\frac{\partial \phi_M}{\partial
z_m}\partial_{\bar z} \phi_M) +
\partial_{\bar z}(\frac{\partial \phi_M}{\partial
z_m}\partial_{z} \phi_M) 
+\frac{\partial \phi_1}{\partial
z_m}\frac{e^\phi}{2}\right)\frac{i dz\wedge d\bar z}{2}. 
\end{equation}
It is easily checked that the only contribution which survives in the
limit $\epsilon \rightarrow 0$ is
\begin{equation}
\frac{\partial S}{\partial z_m} = \lim_{\epsilon
\rightarrow 0}
\int_{X_\epsilon} \frac{e^\phi}{2}\frac{\partial \phi_1}{\partial z_m} 
\frac{i dz\wedge d\bar z}{2}
\end{equation}
which can be computed by using eq.(\ref{eqphiM}) and
$\displaystyle{\frac{\partial \phi_1}{\partial z_m} = -
\frac{g_m}{(z-z_m)}}$
to obtain
\begin{equation}\label{contour}
\frac{\partial S}{\partial z_m} = - i g_m \lim_{\epsilon \rightarrow 0}
\oint_{\gamma_{\epsilon}} \frac{1}{z-z_m}
\partial_z\left(\phi_M+ \alpha_1\phi_B\right) d z.
\end{equation}
We use now  $\phi_M +\alpha_1\phi_B = \phi -\sum_n
g_n\ln|z-z_n|^2-c_0$. 
Both in the elliptic and in the parabolic case we have
\begin{equation}
\partial_z(\phi_M +\alpha_1\phi_B) = -2c_2 -\sum_{n\neq m}
\frac{g_n}{z-z_n}+ \rho_n(\zeta) 
\end{equation}
where $c_2= \beta_m/2g_m$ and the contour integral of
$\rho_n(\zeta)$ in eq.(\ref{contour}) vanishes for
$\epsilon\rightarrow 0$.
Finally we have
\begin{equation}
\frac{\partial S}{\partial z_m} = 
-2\pi \beta_m - 2\pi\sum_{n,n\neq m} \frac{g_m g_n}{z_m-z_n} 
\end{equation}
equivalent to Polyakov conjecture 
\begin{equation}
-\frac{1}{2\pi}\frac{\partial S_P}{\partial z_m} = \beta_m
\end{equation}
due to the
relation (\ref{SPtoS}) between $S$ and $S_P$.

\section*{Appendix 1}

In ref.\cite{licht} Lichtenstein proves among others, the
following result: one can write $\phi = U+v$ where $v$ solves the linear
equation (in modern notation)
\begin{equation}
\Delta_{LB} v = \beta + 4 \pi e^{-\phi_B}\sum_n g_n \delta^2(z-z_n).
\end{equation}
$\Delta_{LB}$ is the Laplace- Beltrami operator on the background
$e^{\phi_B}$ i.e. $4e^{-\phi_B}\partial_z\partial_{\bar z}$
and $\beta(z)$ is a positive function regular in $\mathbb{C}$ except
at the 
singularities in a neighbourhood of which it equals  
in the parabolic case
\begin{equation}
\frac{8}{e^{\phi_B} \zeta\bar\zeta(\ln\zeta\bar\zeta)^2}
\end{equation}
 and in the elliptic case is subject to the
inequality
\begin{equation}
0<\lambda' < \beta (\zeta\bar\zeta)^{-g_n} <\lambda''.
\end{equation}
Moreover
\begin{equation}
\int \beta e^{\phi_B} \frac{idz\wedge d\bar z}{2} = -4\pi\sum_n g_n.
\end{equation}
Most important, $U$ is a bounded function over all $\mathbb{C}$ with
a bound $H$ which depends continuously on the $z_n$ and on the $g_n$.
$U$ obviously solves the equation
\begin{equation}
\Delta_{LB} U + \beta = e^{U+v-\phi_B}
\end{equation}
and thus we can write
\begin{equation}\label{Uinv}
U = \Delta_{LB}^{-1} (e^{U+v-\phi_B} - \beta)
\end{equation}
where $\Delta_{LB}^{-1}$ is the usual Green function $\Delta_{LB}^{-1} =
\frac{1}{4\pi} \ln[(z-z')(\bar z- \bar z')]$.
Eq.(\ref{Uinv}) allows us to put uniform bounds of $\partial_z\phi$ and
$\partial^2_z\phi$ in a domain which excludes the singularities $z_n$.
In fact we have
\begin{equation}
\partial_z U = \frac{1}{4 \pi}\int \frac{1}{z-z'}
[e^{U+v-\phi_B}-\beta](z')
e^{\phi_B(z')} d^2 z'.
\end{equation}
We can write the integral as the sum over a small disk $D_\eta$ of
radius
$\eta$ around $z$ and the remaining of the complex plane. The modulus
of the first term is bounded by
\begin{equation}
e^H\int_{D_\eta}\frac{1}{|z-z'|} e^{v(z')}d^2z'
+\int_{D_\eta}\frac{1}{|z-z'|} |\beta(z')| e^{\phi_B(z')}d^2z'
\end{equation}
and as such bounded when e.g. $z_1$ varies in a small neighborhood
which does not overlap with the other singularities.
The integral over $\mathbb{C}\setminus D_\eta$ is bounded by
\begin{equation}
\frac{1}{\eta}\int
e^{U(z')+v(z')}d^2z'+\frac{1}{\eta}\int
|\beta(z')|e^{\phi_B(z')} d^2z'.
\end{equation}
Both integrals are finite and vary regularly with $z_n$.
Similarly one can put a bound on the second derivative.
\begin{equation}
\partial^2_z U = \frac{1}{4 \pi}\int_{D_\eta } \frac{1}{z-z'}
[e^{U+v-\phi_B} \partial_{z'}(U+v)-\partial_{z'}\beta -\beta
\partial_{z'}\phi_B](z') e^{\phi_B(z')} d^2 z'-
\end{equation}
$$
\frac{1}{4 \pi}\oint_{\partial D_\eta } \frac{1}{z-z'}
[e^{U+v-\phi_B} -\beta](z') e^{\phi_B(z')} \frac{id\bar z'}{2}-
$$
$$
\frac{1}{4 \pi}\int_{\mathbb{C}\setminus D_\eta } \frac{1}{(z-z')^2}
[e^{U+v -\phi_B} -\beta](z') e^{\phi_B(z')} d^2 z'.
$$
The first term is bounded by the sum of the convergent integrals
\begin{equation}
\frac{1}{4\pi}e^H\int_{D_\eta}\frac{1}{|z-z'|} e^{v}
|\partial_{z'}(U+v)| d^2 z'
\end{equation}
$$
\frac{1}{4\pi}\int_{D_\eta}\frac{1}{|z-z'|}|\partial_{z'}\beta -\beta
\partial_{z'}\phi_B|e^{\phi_B(z')}d^2z',
$$
the second is bounded due to uniform boundedness of the integrand on
the contour and the last is bounded similarly as done for the first
derivative.

We recall now that the rational function $Q(z)$ is related to the
analytic 
component of the energy momentum tensor of Liouville theory by
\begin{equation}\label{Qeq}
Q(z) = \sum_n \frac{1-\mu_n^2}{4(z-z_n)^2}+\frac{\beta_n}{2(z-z_n)} =
-\frac{1}{2}[\frac{1}{2}(\partial_z\phi)^2 -\partial_z^2\phi].
\end{equation}
Thus we can extract the $\beta_n$ as contour integrals of the
r.h.s. of eq.(\ref{Qeq}) choosing contours which enclose the
singularities. As $\partial_z\phi$ and $\partial_z^2\phi$ are
uniformly bounded when the $z_n$ vary on small disks of $\mathbb{C}$
which do 
not overlap, we have that on such domain the $\beta_n$ remain bounded.

With regard to the boundedness of
$k_n$ we recall that $\phi$ in the case of elliptic singularities has
the following expansion near $z_n$ 
\begin{equation}
\phi = g_n\ln|z-z_n|^2 -\ln s_n^2 +o(1)
\end{equation}
and $v$
\begin{equation}
v = g_n\ln|z-z_n|^2 +c_n+ o(1)
\end{equation}
where $c_n$ varies continuously with the $z_m$.

Taking into account that $U$ is bounded we have
\begin{equation}
- \ln s^2_n = c_n+ U(z_n)
\end{equation}
and thus $\ln s^2_n$ is bounded when the $z_m$ vary in small non
overlapping disks of $\mathbb{C}$. As according to eq.(\ref{expansion})
\begin{equation}
-\ln s_n^2 =  \ln k_n^2 +\ln(g_n+1)^2+ \ln 8
\end{equation}
the same can be said about the boundedness of $\ln k^2_n$.
Similarly one can deal with the boundedness of $\beta_n$ and of $k_n^2$ as 
functions of the $g_n$ when $g_n$ vary in small domains respecting
Picard's bounds. One can easily extend the proof of the boundedness of
$\beta_n$ and $k^2_n$ to the case of parabolic singularities.

\section*{Appendix 2}

In this appendix we rewrite Polyakov's regularized action in terms of
the field $\phi_M$ and a background field $\phi_B$. We write

\begin{equation}\label{decomposition}
\phi = \phi_M + \phi_1 + \alpha_1 \phi_B
\end{equation}
choosing $\phi_1$
\begin{equation}
\phi_1 = \sum_ng_n\ln(z-z_n)(\bar z-\bar z_n) +c_0.
\end{equation}
$e^{\phi_B}$ is the background conformal factor  describing a surface
with the topology of the sphere. Thus we have
\begin{equation}
-\int e^{-\phi_B} 4\partial_z\partial_{\bar z}\phi_B d\mu \equiv -\int
\Delta_{LB} \phi_B d\mu = -\int 4\partial_z\partial_{\bar z}\phi_B
\frac{idz\wedge d\bar z}{2} = 8\pi.
\end{equation}
This relation fixes the asymptotic behavior of $\phi_B$
\begin{equation}
\phi_B = -2\ln z\bar z + c_B +o(z).
\end{equation}
As a consequence $\phi_M$ solves
\begin{equation}
4\partial_z\partial_{\bar z} \phi_M = e^\phi -\alpha_1
4 \partial_z\partial_{\bar z}\phi_B.
\end{equation}
$\phi_M$ is finite for $z=z_n$ in the case of elliptic singularities, while 
for parabolic singularities it diverges like $\ln \ln^2 (\zeta\bar
\zeta)$; in 
order to have $\phi_M$ regular at infinity we choose $\alpha_1 =
(\sum_n g_n+g_\infty)/2$.   
We shall write $\phi_M(\infty) = c_M$. 
We have
\begin{equation}
\int_{X_\epsilon} \partial_z \phi~ \partial_{\bar z}\phi~
\frac{idz\wedge d\bar z}{2}=
\int (\partial_z \phi_M \partial_{\bar z}\phi_M -2\alpha_1
\phi_M\partial_z\partial_{\bar z }\phi_B)\frac{idz\wedge d\bar z}{
2}+
\end{equation}
$$
\int_{X_\epsilon}\partial_z(\phi_M \partial_{\bar
z}(\phi_1+\alpha_1\phi_B))\frac{idz\wedge d\bar z}{ 2} +
\int_{X_\epsilon}\partial_{\bar z}(\phi_M
\partial_{z}(\phi_1+\alpha_1\phi_B))\frac{idz\wedge d\bar z}{ 2}+
$$
$$
\int_{X_\epsilon}\partial_z(\phi_1+\alpha_1\phi_B) \partial_{\bar
z}(\phi_1+\alpha_1\phi_B)\frac{idz\wedge d\bar z}{ 2}
$$
The second and third integral reduce to
\begin{equation}\label{a2}
-\frac{i}{2}\sum_n g_n \oint_{\gamma_n}\phi_M [\frac{d\bar z}{\bar
 z-\bar
z_n}-\frac{dz}{z-z_n}] -\frac{i}{2} g_\infty \oint_{\gamma_\infty
 }\phi_M
[\frac{d\bar z}{\bar z}-\frac{dz}{z}]
\end{equation}
while the fourth becomes
$$
-\frac{i}{2} \sum_n\oint_{\gamma_n}(\phi_1+\alpha_1\phi_B)\partial_{\bar
z}\phi_1 d\bar z +
$$
\begin{equation}\label{a1}
\frac{i}{2}
\oint_{\gamma_\infty}(\phi_1+\alpha_1\phi_B)\partial_{\bar z}\phi_1
d\bar z  -\int (\phi_1+\alpha_1\phi_B) 
\alpha_1\partial_z\partial_{\bar z}\phi_B
\frac{idz\wedge d\bar z}{ 2}.
\end{equation}
The terms (\ref{a1}) combines with (\ref{a2}) as follows: 
the first two terms in
(\ref{a1}) sum to the integrals in (\ref{a2}) to
give the contour integrals in the original action
(\ref{Sepsilon1}), leaving the term
\begin{equation}
-\alpha_1^2\int\phi_B\partial_z\partial_{\bar z}\phi_B \frac{idz\wedge
d\bar z}{ 2}
\end{equation}
and the terms
\begin{equation}
-\frac{i}{2}\sum_n g_n \oint_{\gamma_n}
(\phi_1+\alpha_1\phi_B) \frac{dz}{z-z_n}
-\frac{i}{2}g_\infty \oint_{\gamma_\infty}
(\phi_1+\alpha_1\phi_B)\frac{dz}{z}-\alpha_1\int\phi_1\partial_z
\partial_{\bar z} \phi_B \frac{idz\wedge d\bar z}{2}.
\end{equation}
These terms do not contain the field $\phi_M$ and can be computed
explicitly; they give rise to the divergence $\pi\sum_n g_n^2 \ln
\epsilon^2 + \pi g^2_\infty \ln\epsilon^2$ which cancel the 
one in the original action (\ref{Sepsilon1}) and to the finite terms
\begin{equation}
\pi \sum_ng_n\sum_{m\neq n} g_m\ln|z_n-z_m|^2 + 2\pi\alpha_1^2 c_B + 4\pi
\alpha_1 c_0. 
\end{equation}
Thus the final form of the action is
\begin{equation}
S_P = \int (\partial_z \phi_M \partial_{\bar z}\phi_M 
+\frac{e^\phi}{2}-2\alpha_1
\phi_M\partial_z\partial_{\bar z }\phi_B)\frac{idz\wedge d\bar z}{
2}+  \pi \sum_ng_n\sum_{m\neq n} g_m\ln|z_n-z_m|^2
\end{equation}
$$
+4\pi\alpha_1 c_0-\alpha_1^2\int\phi_B\partial_z\partial_{\bar z}\phi_B
\frac{idz\wedge d\bar z}{ 2} + 2\pi\alpha_1^2
c_B.
$$

The same procedure holds in the case when one or more singularities
are of the parabolic type yielding the same result. In fact the
kinematic field $\phi_B$ is the
same, $\phi_1$ is obtained by replacing some $g_n$ with $-1$ and the
integral containing the field $\phi_M$ are still absolutely
convergent.

\eject

\end{document}